# New DRIE-Patterned Electrets for Vibration Energy Harvesting

S. Boisseau[1,a], A.-B. Duret[1], J.-J. Chaillout[1], and G. Despesse[1]

[1]CEA, LETI, Minatec Campus, 17 av des martyrs, Grenoble, France

**Abstract.** This paper is about a new manufacturing process aimed at developing stable $SiO_2/Si_3N_4$ patterned electrets using a Deep Reactive Ion Etching (DRIE) step for an application in electret-based Vibration Energy Harvesters (e-VEH). This process consists in forming continuous layers of $SiO_2/Si_3N_4$ electrets in order to limit surface conduction phenomena and is a new way to see the problem of electret patterning. Experimental results prove that patterned electrets charged by a positive corona discharge show excellent stability with high surface charge densities that may reach 5mC/m² on 1.1µm-thick layers, even with fine patterning and harsh temperature conditions (up to 250°C). This paves the way to new e-VEH designs and manufacturing processes.

## 1 Introduction

With the will to increase the number of sensors around us and to respect several economic and environmental constraints, researchers and R&D engineers are looking for new green and unlimited energy sources that will enable to remove batteries or wires and to develop autonomous wireless sensor networks with theoretical unlimited lifetimes. These new sources are based on ambient energy harvesting.

Actually, many ambient sources of energy are available in our environment: the most well-known is probably the energy from the sun that can be turned into electricity thanks to photovoltaic cells. Nevertheless, other sources such as temperature gradients, shocks or vibrations are also usable and are particularly suitable when there is no sun (near machines, engines…) [1]. In this work, we focus on energy harvesting from vibrations.

Many solutions have been developed to convert mechanical energy from vibrations into electricity; they are generally grouped into three categories: piezoelectric devices, electromagnetic devices and finally electrostatic devices. Electrostatic devices, based on a plane capacitor architecture, are the subject of this paper. Specifically, we study here electret-based energy harvesters. To work, electret-based vibration energy harvesters (e-VEH) need electrets.

Electrets are dielectrics capable of keeping a permanent electric charge through time. Unfortunately, due to conduction mechanisms in dielectrics, charges do not stay trapped inside of dielectrics – especially with patterned electrets – which means that, after some time, e-VEH will not work anymore. Keeping charges inside of dielectrics for a long time is a real challenge; we present in this article a new manufacturing process to pattern electrets that has already proven to give $SiO_2/Si_3N_4$ electrets an excellent stability since no charge decay was observed during 1 year.

---

[a] sebastien.boisseau@cea.fr

## 2 Electret-based Vibration Energy Harvesters

We present in this section the concept of vibration energy harvesting using electret-based converters. As we want to harvest small vibrations, we show that electret patterning is necessary in e-VEH.

### 2.1 Vibration Energy Harvesting

Vibration energy harvesters are devices able to turn mechanical energy of vibrations into electricity. The mechanical model presented hereafter can be applied to all these kinds of devices regardless of the principle of conversion (electrostatic, electromagnetic or piezoelectric) and in particular to electrostatic converters using electrets.

In order to harvest energy from vibrations, it is common to use a damped mass-spring system [2]. The mass (m) is suspended in a frame by a spring (k) and damped by forces ($f_{elec}$ and $f_{mec}$). When a vibration occurs $y(t) = Y\sin(\omega t)$, it induces a relative displacement of the mobile mass $x(t) = X\sin(\omega t + \varphi)$ compared to the frame (figure 1). A part of the kinetic energy of the mass is converted into electricity (modeled by an electromechanical force $f_{elec}$), while an other part is lost in friction forces (modeled by $f_{mec}$).

Ambient vibrations are generally low in amplitude; the use of a mass-spring system generates a phenomenon of resonance, amplifying the amplitude of vibrations perceived by the mobile mass and the harvested power. Newton's second law gives the differential equation that rules the movement of the mobile mass (equation 1). Generally, the mechanical friction force can be modelled as a viscous force $f_{mec} = b_m \dot{x}$. The equation of movement can be simplified by using the natural angular frequency $\omega_0 = \sqrt{k/m}$ and the mechanical quality factor $Q_m = m\omega_0/b_m$.

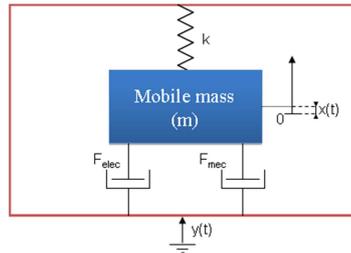

**Fig. 1.** General model of vibration energy harvesters.

This system is ruled by equation (1).

$$m\ddot{x} + f_{meca} + kx + f_{elec} = -m\ddot{y} \Rightarrow \ddot{x} + \frac{\omega_0}{Q_m}\dot{x} + \omega_0^2 x + \frac{f_{elec}}{m} = -\ddot{y} \qquad (1)$$

In e-VEH, the electrostatic force that is needed to turn the mechanical power into electricity is obtained thanks to electrets, inserted into the capacitive structure as the polarization source.

### 2.2 Electrets

Electrets are dielectrics that are able to keep an electric polarization through time. This polarization can be obtained by dipole orientation or by charge injection. Here, we focus on charge injection obtained by a triode corona discharge, which is the quickest way to charge dielectrics. Corona discharge (figure 2) consists in a point-grid-plane structure whose point is subjected to a strong electric field: this leads to the creation of a plasma, made of ions that are projected onto the surface of the sample to charge and whose charges are transferred to the dielectric's surface.

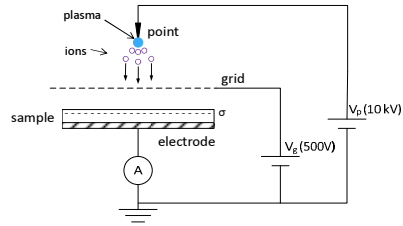

**Fig. 2.** Corona discharge

Charge injection leads to the appearance of a surface potential V that can be easily measured with an electrostatic voltmeter. By assuming that charges are concentrated on the electret's surface: $V = \sigma d/\varepsilon\varepsilon_0$, with ε the electret dielectric permittivity, σ its surface charge density and d its thickness.

Unfortunately, dielectrics are not perfect insulators. As a consequence, some charge conduction phenomena may appear in electrets, and implanted charges can move inside the material or can be compensated by other charges or environmental conditions, and finally disappear. Stability is a key parameter for electrets as the electret-based converter's lifetime is directly linked to the one of the electret.

## 2.3 Electret-based energy converter

Electret-based converters are electrostatic converters, and are therefore based on a capacitive architecture made of two electrodes (figure 3(a)). The presence of the electret induces charges on electrodes and counter-electrodes to respect Gauss's law. Therefore, $Q_i$, the charge on the electret is equal to the sum of $Q_1$ and $Q_2$, where $Q_1$ is the total amount of charges on the electrode and $Q_2$ the total amount of charges on the counter-electrode ($Q_i = Q_1+Q_2$). A relative displacement of the counter-electrode compared to the electret and the electrode, leads to a change in the capacitor geometry and a reorganization of charges between the electrode and the counter-electrode through the load. This induces a current across load R and part of the mechanical energy is then turned into electricity. The equivalent electric model of e-VEH is presented in figure 3(b) and consists in a variable capacitor in series with a voltage source [3].

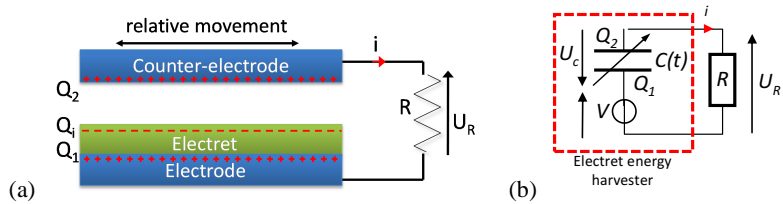

**Fig. 3.** (a) Electret-based electrostatic converter and (b) equivalent electric model

The output power (P) of this converter is directly linked to the electret's surface voltage V and the capacitance variation of the structure dC/dt when it is subjected to vibrations [4]. As a first approximation, it appears that:

$$P \propto V^2 \frac{dC}{dt} \qquad (2)$$

As e-VEH output power is linked to the electret's surface voltage and its lifetime to the electret's lifetime, Surface Potential Decays (SPDs) that consist in measuring the surface voltage of the electret as a function of the time after charging, are the best way to characterize electrets for this application.

To take advantage of a resonance phenomenon, the converter presented in figure 3 is integrated into a mass-spring system introduced in figure 1. However, to increase the variation of capacitance for small displacements, the electrostatic converter has to be patterned as presented in figure 4 (a) and (b) [5].

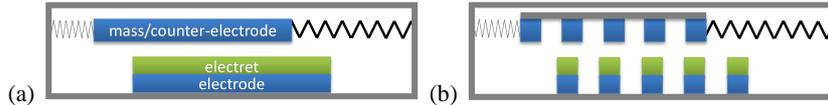

**Fig. 4.** (a) Non-patterned e-VEH and (b) Multibumps e-VEH using patterned electrets enabling to have a high variation of capacitance whatever the vibrations' amplitudes.

Unfortunately, electret patterning is not an easy task as it generally leads to a weak stability of electrets (and especially for $SiO_2$-based electrets). Therefore, we have developed a new manufacturing process aimed at making stable patterned electrets using the well-known $SiO_2/Si_3N_4$ structure.

## 3 $SiO_2/Si_3N_4$ electrets, new patterning and manufacturing process

We present in this section our new patterned electrets' manufacturing process that enables to keep charges for a long time.

### 3.1 $SiO_2/Si_3N_4$ full sheet electrets

Actually, $SiO_2/Si_3N_4$ electrets are known for a long time for being good charge keepers provided that they have received a good thermal treatment and a good surface treatment [6-8]. We have verified the good stability of these electrets. $SiO_2/Si_3N_4$ electrets are obtained from a standard 200mm p-doped silicon wafer. The substrate undergoes a thermal oxidization to form a 1µm-thick $SiO_2$ layer. After a deoxidization to remove the $SiO_2$ layer on the rear face, a 100nm-thick LPCVD layer is deposited on the front face.

Then, after a thermal treatment at 450°C during 2 hours in a protective atmosphere ($N_2$), a HMDS layer (10nm) is deposited on the front face. The final electret structure is presented in figure 5(a).

Samples are then charged thanks to a standard corona discharge. Figure 5(b) presents an example of Surface Potential Decays (SPD) for two different samples charged at 100V. Samples that have not received the thermal and surface treatments lose their charges quickly; samples have lost more than 50% of their initial charge after 1 month. On the opposite, $SiO_2/Si_3N_4$ electrets that have received surface and thermal treatments show an excellent stability and have kept their initial surface voltage even after 600 days.

These results prove (i) the interest of thermal and surface treatments to enable $SiO_2/Si_3N_4$ electrets to keep their charges for a long time and above all (ii) the excellent stability of these electrets through time.

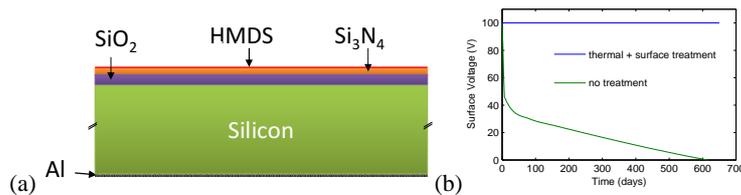

**Fig. 5.** (a) Full sheet $SiO_2/Si_3N_4$ electret and (b) stability through time

### 3.2 New electret patterning solution

As presented in section 2, electret patterning is necessary to develop viable e-VEH. While it is quite easy to pattern organic electrets [4, 9], the same action on inorganic materials is much more difficult as it generally leads to an important charge decay through time and therefore to a limited lifetime of the e-VEH [10, 11].

Nevertheless, inorganic materials such as $SiO_2$ can be really interesting for energy harvesting as they are able to keep a more important surface charge density (up to 12mC/m² [12, 13]) than organic materials which rarely exceed 2mC/m² [14-19].

Naturally, solutions have been developed to pattern $SiO_2$-based structures and especially by IMEC [6, 20], Sanyo and the University of Tokyo [21] and by Tohoku University [11]. Unfortunately, many of these structures cannot theoretically induce a large capacitance variation of the e-VEH. Indeed the capacitance is more or less constant during the relative displacement of the mobile mass compared to the fixed part as the electret surface is more or less flat: e-VEH output power is therefore highly limited.

To increase capacitance variation, one solution would consist in etching the electret and the wafer presented in figure 5(a) to decrease e-VEH minimal capacitance as presented in figure 6(a).

Unfortunately, this obvious patterning does not work for two main reasons:
(i) it is hard to charge these electrets with a corona discharge as ions and charges emitted by corona discharge take the shortest way to the ground and go directly on the silicon without passing through dielectrics.
(ii) it leads to a break of the electret layer continuity that increases charge decay.

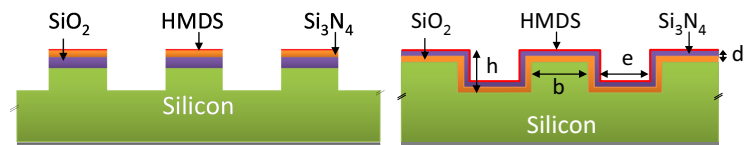

**Figure 6.** (a) An obvious patterning that does not work and (b) Our new patterning

In order to get patterned electrets' stabilities equivalent to full sheet electrets' ones, we have searched for a way to keep a continuous electret layer while maintaining patterning (figure 6(b)). We have rethought the problem of electret patterning: instead of patterning the electret, we have patterned the silicon wafer thanks to a Deep Reactive Ion Etching in order to etch deep trenches (h>100µm). This structure enables to reach a high capacitance variation which is directly linked to the trenches' deepness (figure 7).

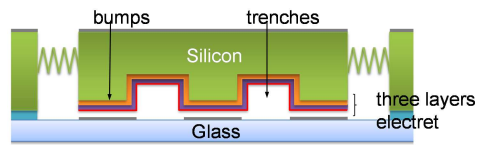

**Figure 7.** Patterned electrets inserted into the electrostatic VEH

### 3.3 Manufacturing process

The fabrication process of these patterned electrets is similar to the one of full sheet electrets in order to keep equivalent behaviours and above all equivalent stabilities.

The main difference between the two processes is the DRIE that is used to geometrically pattern the electret. The main manufacturing steps are presented in figure 8. The process starts with a standard p-doped silicon wafer (a). After a lithography step, silicon wafer is etched by DRIE (b) and cleaned. Wafers are then oxidized to form a 1µm-thick $SiO_2$ layer (c). $SiO_2$ layer on the rear face is then removed by HF while front face is protected by a resin. A 100nm-thick LPCVD $Si_3N_4$ is then deposited on the front face (d). Wafers receive a thermal treatment (450°C during 2 hours into $N_2$) and a surface treatment (vapour HMDS) (e). Dielectric layers are then charged by a standard corona discharge to turn them into electrets.

After charging, samples are stored in a box to protect them from air ions. Relative Humidity (RH) and temperature are also controlled (RH=20%, T=25°C) as they have generally both a large impact on charge stability.

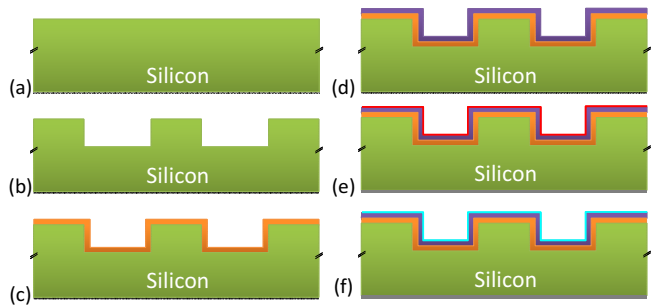
**Figure 8.** Fabrication process of DRIE-patterned electrets

Manufacturing results are presented in figure 9. SEM images in Figure 9 (a, b) show the patterning of the samples and the different constitutive layers. It is interesting to note the continuity of the electret layer even on the right angle in figure 9(c). Patterning is also presented in figure 9(d) on a photo.

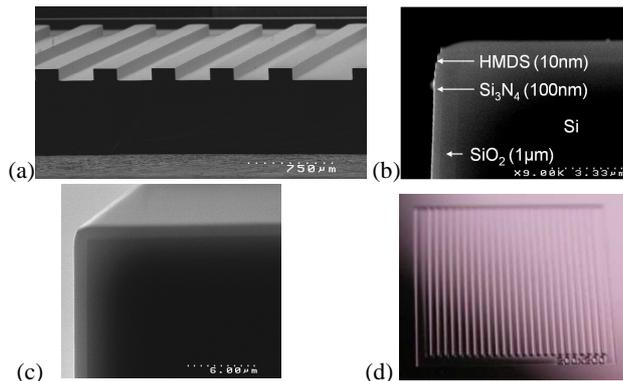
**Figure 9.** DRIE-patterned electrets

Due to the geometric shapes of these patterned electrets, it is not possible to use Kelvin probe force microscopy (KPFM) to get the exact charge repartition on electrets (on bumps, on trenches…). However, Scanning Electron Microscopy (SEM) enables to get an approximate vision of charge localization by highlighting charged areas (that are light on the screen) from non-charged areas (that are dark on the screen).

This experiment has been performed on a DRIE-patterned electret that has been charged only on one half of its surface. Figures 10(a) (charged area) and (b) (non-charged area) show SEM results and confirm that charges seem to be uniformly present on charged electrets.

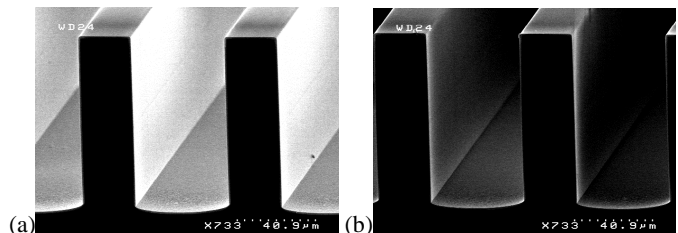
**Figure 10.** DRIE-patterned electrets observed with SEM (a) charged electret and (b) non-charged electret

Therefore, we have managed to get continuous electret layers thanks to this process. We have then tested their stability through time.

### 3.3 Experimental results – Electret stability vs time

Experiments have proven that, with our process, patterned electrets have an excellent stability even with a surface charge density up to 5mC/m² (150V of surface potential on a three-layer 1.1µm-thick electret).

Some of our results are presented in figure 11 for various dimensions (e, b, h) (figure 6(b)) and various initial surface voltages ($V_0$) (obtained from various grid voltages).

Figure 11(a) presents the SPD of a patterned electret (e=25µm, b=25µm) with an initial surface voltage of 68V (3mC/m²). This sample lost less than 1% of its initial charge in 200 days.

We noted that some decays appear when exceeding ~4mC/m², as presented in figure 11(b). The samples (e=100µm, b=100µm, h=100µm) were charged with various initial surface voltages, from 50V to 220V. A first charge decay during the first days can be observed for the samples that have a surface voltage higher than 150V. After 20 days, the surface voltage is stabilized.

We have also tested the effect of a post-charging thermal treatment on the electret stability. A patterned electret (figure 11(c)) was charged with an initial surface voltage of 100V. As expected, no charge decay was observed during the first 20 days. The sample then received a thermal treatment of 250°C during 2H which did not affect its stability.

Finally, we have tested the impact of the patterning size on 100V electrets (figure 11(d)). This chart represents the normalized surface voltage $V/V_0$ (with $V_0$=100V) for various patterning sizes in microns after 100 days. It proves that for 100V electrets, the patterning size slightly impacts the charge stability of $SiO_2/Si_3N_4$ patterned electrets.

These results prove the stability of our patterned electrets and their validity for an application in MEMS VEH.

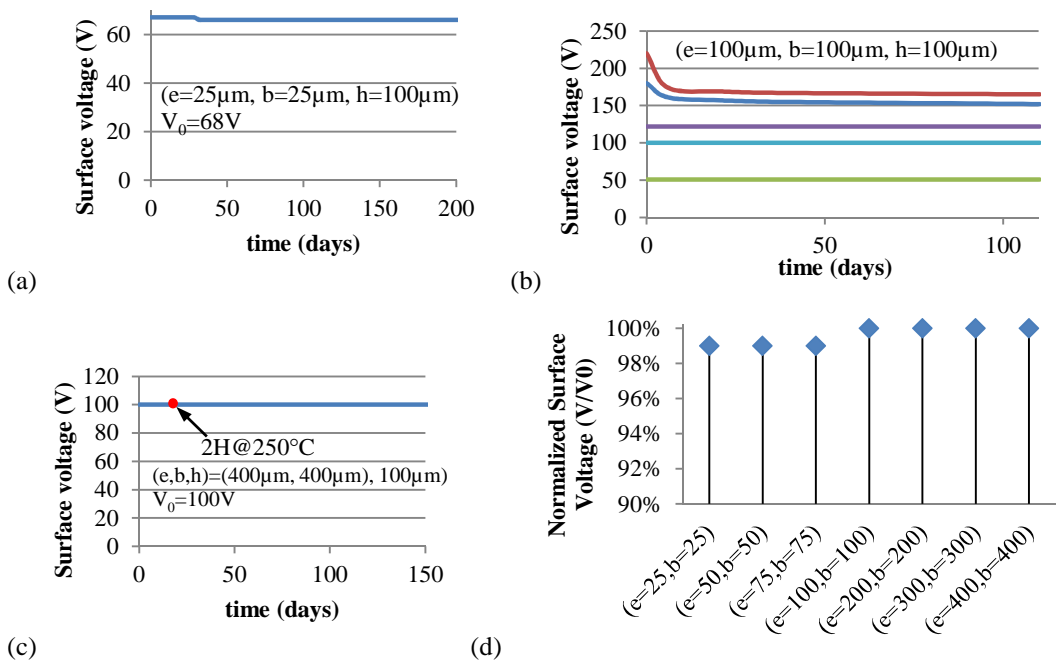

**Figure 11.** Examples of surface potential decays of patterned electrets

We have proven that it is possible to manufacture stable patterned electrets down to 25µmx25µm. We were able to have a good surface charge density that may reach up to ~5mC/m², and we expect to increase this value in the next months.

## 4 Conclusions

We have developed new patterned electrets enabling to induce large capacitance variations on e-VEH, stable through time even when working with high temperatures (up to 250°C). This new patterning solution also presents two great advantages. The first one is the charging of the sample's whole surface with the same surface voltage, limiting conduction surface phenomena between charged and non-charged areas. The second advantage is an easy manufacturing process with only 5 key steps (lithography, DRIE etching, thermal oxidization, $Si_3N_4$ LPCVD and metallization of the rear surface).

These electrets are currently being applied to our MEMS electrostatic VEH. Studies are also being carried out to apply these new electrets to microfluidics, in order to increase surface hydrophobicity by using the double effect: surface patterning and surface voltage.